\documentclass[aps,twocolumn,prx,amsmath,floatfix,superscriptaddress,longbibliography]{revtex4-1}

\usepackage{microtype}
\usepackage{wasysym}

\usepackage[varg]{txfonts}

\usepackage{amssymb}
\usepackage{graphicx}
\usepackage[percent]{overpic}
\usepackage{mathrsfs}
\usepackage{bm}
\usepackage{braket}

\usepackage[colorlinks,bookmarks=false,citecolor=blue,linkcolor=red,urlcolor=blue]{hyperref}

\usepackage{color}

\definecolor{cL}{RGB}{59, 83, 140}
\definecolor{cM}{RGB}{33, 145, 141}
\definecolor{cH}{RGB}{95, 202, 98}
\definecolor{cG}{RGB}{204,204,204}

\begin{document}

\title{Entanglement of mid-spectrum eigenstates of chaotic many-body systems --- reasons for deviation from random ensembles}

\newcommand{\mpipks}{Max-Planck-Institut f\"{u}r Physik komplexer Systeme, D-01187 Dresden, Germany}

\author{Masudul Haque}
\affiliation{Department of Theoretical Physics, Maynooth University, Co.\ Kildare, Ireland}
\affiliation{\mpipks}
\affiliation{Institut f\"{u}r Theoretische Physik, Technische Universit\"{a}t Dresden, D-01062 Dresden, Germany}

\author{Paul A. McClarty}
\affiliation{\mpipks}

\author{Ivan M. Khaymovich}
\affiliation{\mpipks}
\affiliation{Institute for Physics of Microstructures, Russian Academy of Sciences,
603950 Nizhny Novgorod, GSP-105, Russia}

\begin{abstract}

 Eigenstates of local many-body interacting systems that are far from spectral edges are thought to
 be ergodic and close to being random states. This is consistent with the eigenstate
 thermalization hypothesis and volume-law scaling of entanglement. We point out that systematic
 departures from complete randomness are generically present in mid-spectrum eigenstates, and focus
 on the departure of the entanglement entropy from the random-state prediction. We show that the
 departure is (partly) due to spatial correlations and due to orthogonality to the eigenstates at
 the spectral edge, which imposes structure on the mid-spectrum eigenstates.

\end{abstract}

\date{\today}

\maketitle

\section{Introduction}

Ergodicity and equilibration in the quantum realm remain imperfectly understood, and the
characterization of quantum ergodicity is now an active research front. One view is that quantum
ergodicity corresponds to eigenstates of many-body systems being effectively random. This idea is
closely connected to the eigenstate thermalization hypothesis (ETH)
\cite{Deutsch_PRA1991,Srednicki_PRE1994, Rigol_Nature2008,
  Reimann_NJP2015,DAlesio_Polkovnikov_Rigol_AdvPhys2016, Deutsch_RepProgPhys2018,
  Mori_Ikeda_Ueda_thermalizationreview_JPB2018}, and to ideas loosely known as (canonical)
typicality \cite{Tasaki_PRL1998, Gemmer_Otte_Mahler_2ndlaw_PRL2001, Goldstein_Lebowitz_PRL2006,
  PopescuShortLindner_NatPhys2006, Sugita_2007_basis_of_quantstatmech, Reimann_typicality_PRL2007,
  Gemmer_QuantThermo_book2009, Goldstein_Lebowitz_Tumulka_EPJH2010, vonNeumann_translation_EPJH2010,
  Sugiura_Shimizu_canonical_PRL2013, Tasaki_JStatPhys2016_typicality,
  Mori_Ikeda_Ueda_thermalizationreview_JPB2018}.  For a non-integrable (chaotic) many-body
Hamiltonian $H$, it is expected that a state $\ket{\psi_R}$ with independent Gaussian random
coefficients should be a good model for infinite-temperature eigenstates, while eigenstates at
energy corresponding to temperature $1/\beta$ should be well-described by
$e^{-\frac{1}{2}\beta{H}}\ket{\psi_R}$ \cite{White_METTS_PRL2009, Sugiura_Shimizu_TPQstates_PRL2012,
  Sugiura_Shimizu_canonical_PRL2013, Elsayed_Fine_PureQuantumStates_PRL2013,
  Watanabe_Sugiura_NatComm2018_VolumeLaw}.  This expectation is mirrored by the behavior of the
entanglement entropy (EE) in eigenstates of many-body systems with finite Hilbert spaces: At the
spectral edges, EE is low (``area law'') \cite{Eisert_Plenio_RMP2010_AreaLaws,
  Amico_Fazio_RMP2010_EntanglementReview}, while in the infinite-temperature (mid-spectrum) regime,
the eigenstates have EE close to the value expected for random states. As a result, for chaotic
many-body systems, the scatter plot of EE versus eigenenergy takes the shape of an arch or rainbow,
by now familiar from many numerical examples \cite{Beugeling_entanglement_JSM2015,
  garrison2018SingleEigenstate, Turner_Abanin_Serbyn_scarred_PRB2018,
  Moudgalya_Regnault_Bernevig_AKLT_entngl_PRB2018, Mudry_Castelnovo_Chamon_Neupert_PRRes2019,
  Sen_Sen_Sengupta_Floquetscars_PRB2020, Sen_Sen_Sengupta_drivenRydberg_2020,
  Iadecola_Schecter_PRB2020, Mark_Lin_Motrunich_PRB2020, Shibata_Katsura_OnsagerScars_PRL2020,
  Pichler_Lukin_Ho_PRB2020_xy, Mark_Motrunich_2020_etapairing, McClarty_Haque_Sen_Richter_2020}.

In this work, we consider the bipartite entanglement entropy of mid-spectrum eigenstates. For
definiteness, we focus on spin-$1/2$ chains with $L$ sites with all symmetries broken, so that the
Hilbert space is $D=2^L$, and consider the entanglement between two subsystems ($A$, $B$) of equal
size.  In this case, the random states have an average EE well-approximated by the
Page formula \cite{page1993, SM}, $S_{\rm Page} = \log D_{A}-\frac{1}{2}$, where $D_{A}=2^{L/2}$ is
the size of the reduced Hilbert space of the $A$ subsystem.
Although the mid-spectrum eigenstates are expected to be random, numerically calculated mid-spectrum
entanglement in finite-size many-body systems~--~both in the existing literature
\cite{Beugeling_entanglement_JSM2015, Balents_SYK_entanglement_PRB2018,
  LeBlond_Vidmar_Rigol_PRE2019_integrable} and in this work~--~systematically fall below the Page
value.  The deviation decreases more slowly with system size than the width of the state-to-state
fluctuations of EEs, which means that the deviation is significant at {\it any} finite size. To the
best of our knowledge, the origin of this subtle, systematic and seemingly universal effect has not
been addressed so far. In this work, we present a study of this discrepancy, uncovering the ways in
which mid-spectrum eigenstates deviate from random states.

\begin{figure}[tbp]
\centering
\hspace{-.5cm}
\centering
\includegraphics[width=\columnwidth]{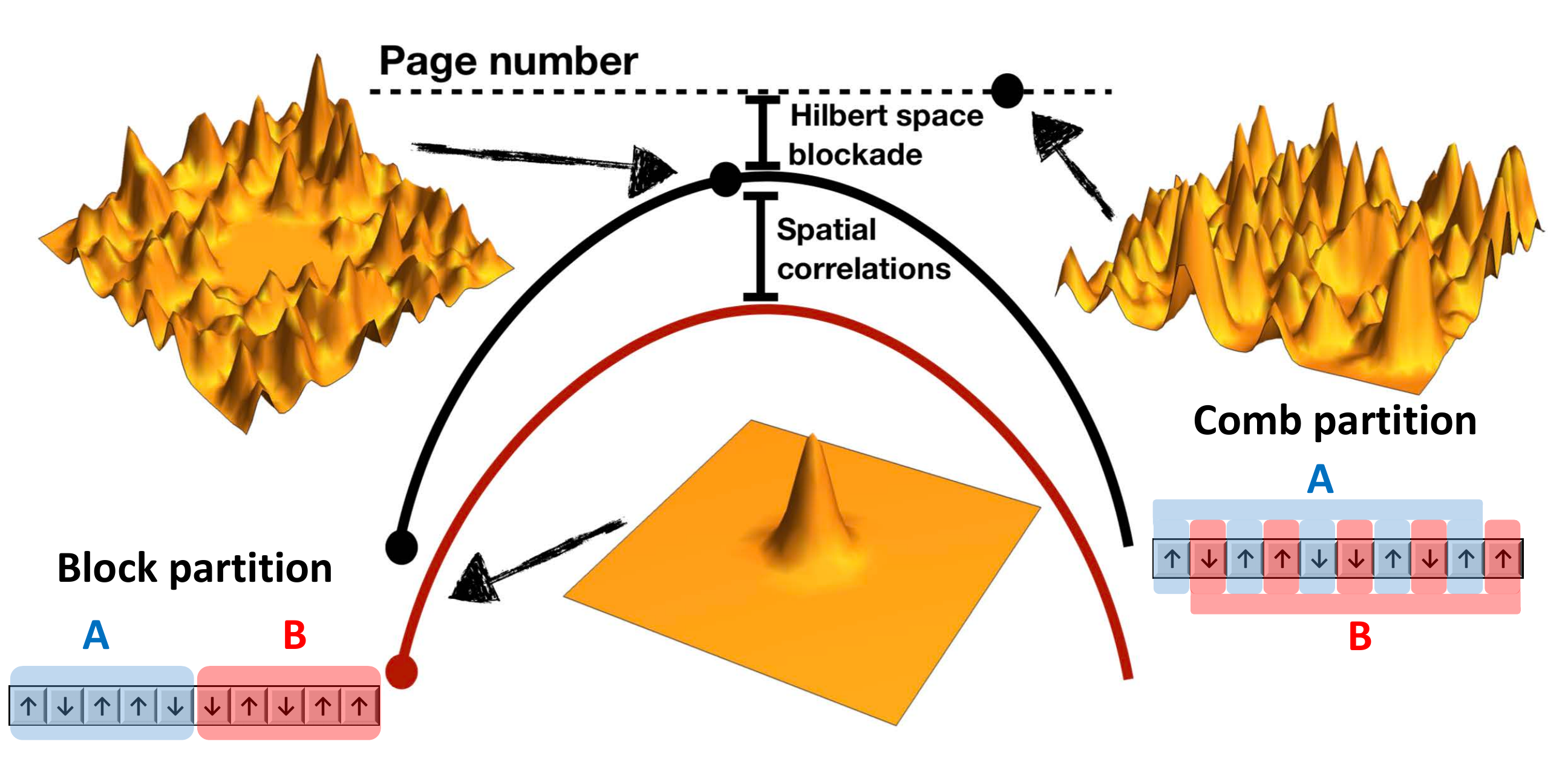}
\caption{Schematic, summarizing main ideas.  Lower rainbow shows block EE against
  eigenenergies. Upper rainbow is EE between ``comb'' partitions. Partitions are shown in
  the left and right insets, respectively.  The difference between the top of the two rainbows is
  attributed to residual spatial correlation in mid-spectrum eigenstates.  The difference between
  the Page value (dotted horizontal line) and the mid-spectrum comb entanglement is attributed to
  the ``orthogonality blockade'' effect --- orthogonality to the special spectral edge states.  3D
  plots are cartoons of distributions of eigenstate intensities, using a 2D space to visualize the
  Hilbert space.  The three cases correspond to low-entanglement states typical of spectral edges
  (middle), fully random or `ergodic' states (right), and  mid-spectrum states (left). The
  latter demonstrates a depletion of weight in parts of the Hilbert space where the spectral edges
  have large weight.
\label{fig:1}
}
\end{figure}

We find that the {\it locality} of the Hamiltonian leads to {\it spatial correlations} persisting in
mid-spectrum eigenstates of any finite system; we demonstrate this through the mutual information
between sites.  The mid-spectrum eigenstates thus differ in an important manner from random states.
The presence of spatial correlations manifests itself strongly in the entanglement between spatially
connected blocks ({\it block} bipartition) -- a partitioning which is naturally sensitive to spatial
correlations in eigenstates.  We show that the departure of mid-spectrum entanglement from the Page
value is smaller for {\it comb} partitions that are, of all bipartitions, the least sensitive to
spatial variations of correlations.
Nevertheless, even the comb entanglements depart from the Page value.  We argue that the reason for
the departure from Page value of the mid-spectrum eigenstates is their orthogonality to the
eigenstates at spectral edges. Orthogonality forces the mid-spectrum eigenstates to live in an
effectively lower-dimensional Hilbert space: part of the physical Hilbert space is blocked off. This
{\it Hilbert space blockade} phenomenon manifests itself in the eigenstate coefficient distribution
as an enhanced weight around zero~\cite{Beugeling_coefficients_PRE2018,LuitzBarlev_PRL16,
  Baecker2019,Luitz_Khaymovich_BarLev_multifrac_SciPost2020}.  The orthogonality blockade effect
exists for any Hamiltonian, local or not. However, for local Hamiltonians, eigenstates at the
spectral edges have area law entanglement and strong spatial correlations. The blockade effect then
forces mid-spectrum states to have the observed spatial correlations.  The departure from the Page
value for comb partitions is thus due to a correlated Hilbert space blockade, such that certain
types of configurations in the Hilbert space are blocked from appearing in the mid-spectrum
eigenstates.
This scenario is illustrated in Fig.~\ref{fig:1} and elaborated in the
rest of this paper.

\section{Model}

We focus on the spin-$\frac{1}{2}$ chain, with couplings between sites $i$, $j$
having the XYZ form
$h_{i,j}[\eta,\Delta] = (1-\eta) S^x_iS^x_{j}+ (1+\eta) S^y_{i}S^y_{j} +\Delta S^z_{i}S^z_{j}$.
The nearest-neighbor version of this model is integrable through the algebraic Bethe ansatz
\cite{McCoy_AdvancedStatMech_book2009}. Since our focus is on non-integrable systems, we add
next-nearest neighbor couplings $h_{j,j+2}$ and/or magnetic fields:
\begin{multline}\label{eq:model}
H = J_1 \sum_{j=1}^{L-1} h_{j,j+1}[\eta_1,\Delta_1]
 + J_2 \sum_{j=2}^{L-2} h_{j,j+2}[\eta_2,\Delta_2]
\\ + h_z \sum_{j=1}^{L} (1-\tfrac{1}{2}\delta_{jL}) S^z_{j}
+ h_x \sum_{j=1}^{L} (1-\tfrac{1}{2}\delta_{j1}) S^x_{j} .
\end{multline}
The $h_x$ term breaks the parity of total-$S^z$. In addition, the $J_2$, $h_z$ and $h_x$ terms are
each tweaked at one edge of the chain so that reflection symmetry is broken. Unless otherwise
specified, we present data for $J_\alpha=1$, $\eta_\alpha=0.5$, $\Delta_\alpha=0.9$, $h_z=0.8$,
$h_x=0.2$.
For parameters that we used, the level spacing statistics of the model
is consistent with that of the Gaussian orthogonal ensemble (GOE),
indicating chaotic behavior.


\begin{figure}[tbp]
\centering
\hspace{-.5cm}
\centering
\includegraphics[width=\columnwidth]{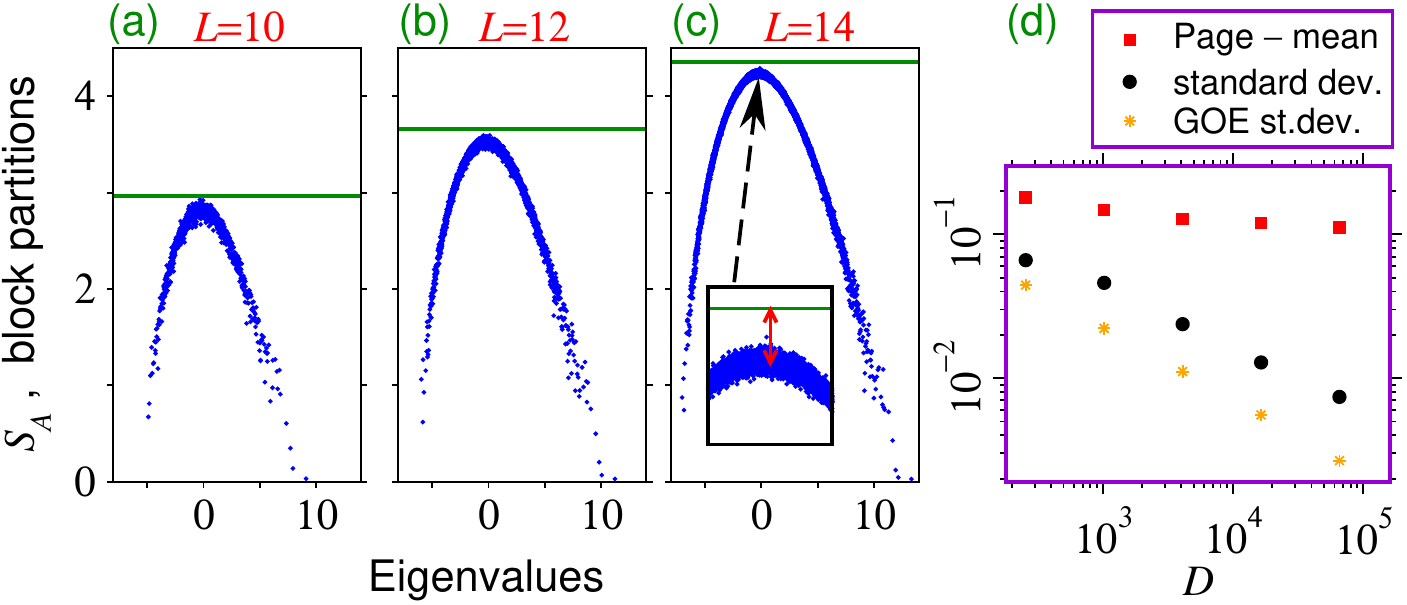}
\caption{EE for block bipartition.  Parameters listed in text.  (a-c)~EE versus energy for different
  system sizes.  Horizontal line: Page value.  (d)~Statistics of mid-spectrum eigenstates ($D/16$
  states nearest to rainbow peak), for $L=$8,10,12,14,16.  The standard deviation has
  similar scaling $\sim D^{-1/2}$ to the GOE case.  The departure from the Page value
  falls off much more slowly, possibly even saturating.
\label{fig:2}
}
\end{figure}

\section{Entanglement for block partitions}

In Fig.~\ref{fig:2} we consider ``block'' bipartitioning, i.e., the $A$ ($B$) bipartition is the
left (right) half of the chain.  The spectral edges and mid-spectrum states scale differently
($\sim{L^0}$ vs $\sim{L^1}$), resulting in the rainbow/arch shape, Fig.~\ref{fig:2}(a-c).  The
largest EE values are close to the EE values of random states of the same Hilbert space size, whose
average is here the Page value, $S_{\text{Page}}=\frac{L}{2}\ln2-\frac{1}{2}$, because we have
chosen a spin-$\frac{1}{2}$ system with no symmetries.  The EE being close to the Page value
indicates that the mid-spectrum eigenstates are close to being ``random'' or
``infinite-temperature''.  Accordingly, the width of the distribution of mid-spectrum EE values is
expected to decrease as $\sim{D}^{-1/2}$, like eigenstates of GOE/GUE matrices
\cite{Vivo_Oshanin_PRE2016}.  Fig.~\ref{fig:2}~(d) shows that the mid-spectrum EE widths are larger
than corresponding GOE values, but are consistent with $\sim{D}^{-1/2}$ behavior.

The feature we focus on in this paper is the systematic departure from the Page value even in the
middle of the spectrum (Fig.~\ref{fig:2}~(c) inset).  Fig.~\ref{fig:2}~(d) shows how the departure
of the mean mid-spectrum EE from the Page value scales with system size.  There is some ambiguity in
how to choose the ``mid-spectrum'' states.  Fig.~\ref{fig:2}~(d) uses the 1/16th eigenstates closest
to the top of the rainbow, but our observations are insensitive to the exact procedure~\cite{SM}.
The departure decreases with the system size \emph{remarkably slowly}.  In fact, the data does not
rule out saturation, i.e., nonzero departure in the thermodynamic limit.  The departure is certainly
much larger than the width, which decreases much faster, $\sim{D}^{-1/2}$.  Thus, at \emph{any}
system size, the Page value lies outside the distribution of EE values.  In this sense, the
departure is not ``merely a finite size effect.''



\section{Spatial Correlations and comb entanglement}

\begin{figure}[tbp]
\centering
\hspace{-.5cm}
\centering
\includegraphics[width=1.0\columnwidth]{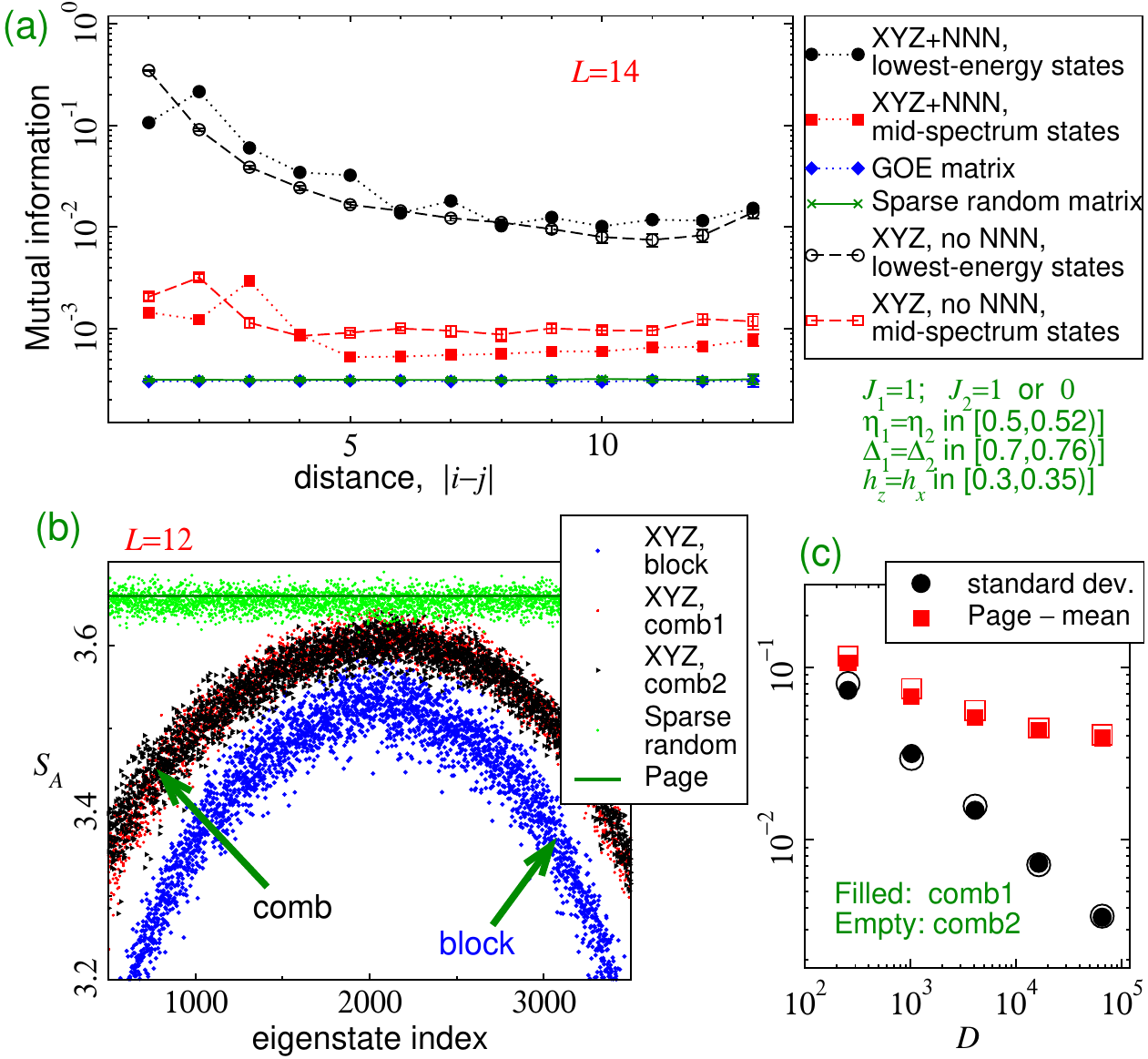}
\caption{Spatial correlations and comb EE.
(a)~Mutual information vs.\ distance.  Low-energy and mid-spectrum eigenstates of XYZ Hamiltonians,
compared with eigenstates of full (GOE) and sparse random
matrices.  Statistics gathered from 32 eigenstates of each of 40 XYZ Hamiltonians, with parameters drawn from ranges
shown.
(b)~EE for block and comb bipartitions.
%
%
(c)~Scaling of mid-spectrum comb EE (both types, filled/open symbols), similar to
Fig.~\ref{fig:2}(d).  The departure from the Page value falls much slower than the width.
\label{fig:3}
}
\end{figure}

To uncover the reason for the departure from
random-matrix behavior, we first appeal to the best-known case of such departures, namely the
spectral edges, for which the volume law of entanglement is violated
\cite{Eisert_Plenio_RMP2010_AreaLaws, Amico_Fazio_RMP2010_EntanglementReview}.  The origin of area
law EE is the spatial locality of the Hamiltonian.  This causes correlations to decay rapidly with
distance, and ensures that the block entanglement receives its largest contribution from the
boundary region.
%
%
We ask whether some degree of locality, in this sense, also exists in the mid-spectrum eigenstates.
We quantify correlations via the quantum mutual information
\begin{equation}
  I(i,j) \equiv S_{\rm [i]}+S_{\rm [j]}-S_{\rm [i\cup j]}
\end{equation}
for pairs of spins $i$ and $j$, as in
\cite{DeTomasi_Bera_Bardarson_Pollmann_MutInf_PRL2017, DeTomasi_algebraicMBL_PRB2019}.
Fig.~\ref{fig:3}(a) shows $I(i,j)$ against the distance $|i-j|$, for mid-spectrum eigenstates,
low-energy eigenstates and random states.  The distance-dependence in mid-spectrum eigenstates is
much less pronounced than in low-energy eigenstates, as expected.  However, there exists an
unmistakable dominance of small-distance correlations.  Thus, we identify distance-dependence as one
reason for the departure of mid-spectrum EE from the Page value.

To `correct' for this effect, we consider `comb' bipartitions \cite{Keating_Mezzadri_comb_PRA2006,
  Zanardi_comb_NJP2006, Igloi_Peschel_disjoint_EPL2010, Rossignoli_comb_PRA2011,
  Magan_Vandoren_sublatticeEE_PRB2017} of two types: `comb1' partitions spins as $ABABAB...$, i.e.,
it is a sublattice partition, while `comb2' partitions the spins as $AABBAABB...$.  For comb2
partitioning, there is some ambiguity when $L$ is not divisible by $4$; we resolve this by  setting the
last two sites to be in partitions $A$ and $B$.  For example, for $L=10$, the comb2 partitioning is
$AABBAABBAB$.

For such bipartitions, the EE should be insensitive to large-scale distance-dependence of
correlations, as two nearby points are as likely to be in different partitions as two faraway
points.  (The boundary and bulk of partitions are not spatially separated.)  Fig.~\ref{fig:3}(b)
shows that the mid-spectrum comb EE is indeed closer to the Page value than the block EE.  There is
little difference between comb1 and comb2, which supports the idea that this reduction of the
departure is due to removal of the effect of distance-dependence.  For comb bipartitioning, there is
no notion of `area law', so that the EE for spectral edges scale as $\sim{L}$ instead of
$\sim{L^0}$.  Nevertheless, the comb EE's at the spectral edges are significantly smaller than the
mid-spectrum ones, despite having the same scaling.  Thus the comb EE's are also arranged in an
arch/rainbow shape~\cite{SM}.

Remarkably, even for comb partitions, the departure from the Page value remains much larger than the
width of mid-spectrum EE distributions.  As in the block case, the comb EE departure decreases far
slower than the width, Fig.~\ref{fig:3}(c).  (The available data would even be consistent with a
\emph{saturation} of the departure in the $L\to\infty$ limit, as opposed to a slow decrease.)  Thus
the departure is a visible effect at any finite size, also for partitions which (unlike block
partitions) do not select for locality effects.  We are thus forced to look for additional
mechanisms -- beyond `locality' as discussed above -- for the departure from the random-state
behavior.

\section{Non-explanations}

\emph{Sparsity is not responsible ---}
One possible source of the difference between random states and the eigenstates of local many-body
Hamiltonians is that such Hamiltonians are generally sparse matrices in common basis choices.  To
examine the consequence of sparsity, in Fig.~\ref{fig:3}(a,b) we include results (mutual
information, EE) for the eigenstates of matrices with sparsity close to the physical (XYZ)
Hamiltonian, and nonzero elements drawn from a Gaussian distribution.  We find only very slight
differences from eigenstates of the usual (full) GOE ensemble --- in Fig.~\ref{fig:3}(a) the mutual
information values for the GOE case and the sparse random case are very slightly offset from each other
(offset barely visible), while in Fig.~\ref{fig:3}(b) the EE values for the sparse random matrix
have a distribution whose center is only very slightly lower than the page value.

Thus, sparsity is not a significant factor in the departure from the Page value.

\emph{Effect of finite measurement window ---} To obtain sufficient statistics for the average and
width of mid-spectrum entanglements, we use the EE values within some energy window,
$E_{\max}-\Delta E/2<E<E_{\max}+\Delta E/2$, containing the energy $E_{\rm max}$ where the EE is
maximal.  Approximating the EE to be a smooth function of energy, $S(E)$, the average EE within an
energy window $\Delta{E}$ is obtained by Taylor expansion to be smaller than the maximum by the
amount $S_{\rm Taylor}=\frac{1}{24}|S''(E_{\rm max})|(\Delta{E})^2$.  We are of course interested in
the limit $\Delta{E}\to0$, where this effect plays no role.  To ensure that we have reached this
limit, we have carefully checked that our extracted values of departure are independent of the
$\Delta{E}$ value used numerically, and also that the Taylor correction term is orders of magnitude
smaller than the departure, for the values of $\Delta{E}$ used numerically~\cite{SM}.

In addition, from Fig.~\ref{fig:2}(c) inset and from Fig.~\ref{fig:3}(b), it is visually obvious
that the top of the rainbow itself deviates from the Page value, and that the effect is not due to
averaging over a finite energy window.

\section{Orthogonality and blockade}

We now introduce a framework for discussing the deviation of mid-spectrum states from randomness
(full ergodicity).  Eigenstates at the spectral edges are well-known to be special -- they have
area-law instead of volume-law entanglement, and this is reflected in the local structure of
correlations.  These eigenstates may be seen as occupying a specific tiny part of the Hilbert space
which promotes the special features.  Because mid-spectrum eigenstates need to be orthogonal to
these special states, they are forced to \emph{exclude} that part of the Hilbert space.  Thus
mid-spectrum eigenstates are distributed in a large fraction of, but not the complete, Hilbert
space: part of the Hilbert space is blocked off.

This blockade phenomenon is illustrated in the cartoons of Fig.~\ref{fig:1}.  If we use real-space
configurations as the basis, then these cartoons over-simplify in showing the low-energy states as
having exactly zero coefficients for most basis states.  In reality, the low-energy eigenstates are
not completely localized in configuration space --- they have low but nonzero entanglement.  In
other words, their participation ratios \cite{SM} are much smaller than the random state value of
$D/3$, but are still much larger than $1$ \cite{Beugeling_entanglement_JSM2015}.

\begin{figure}[tbp]
\centering
\hspace{-.5cm}
\centering
\includegraphics[width=\columnwidth]{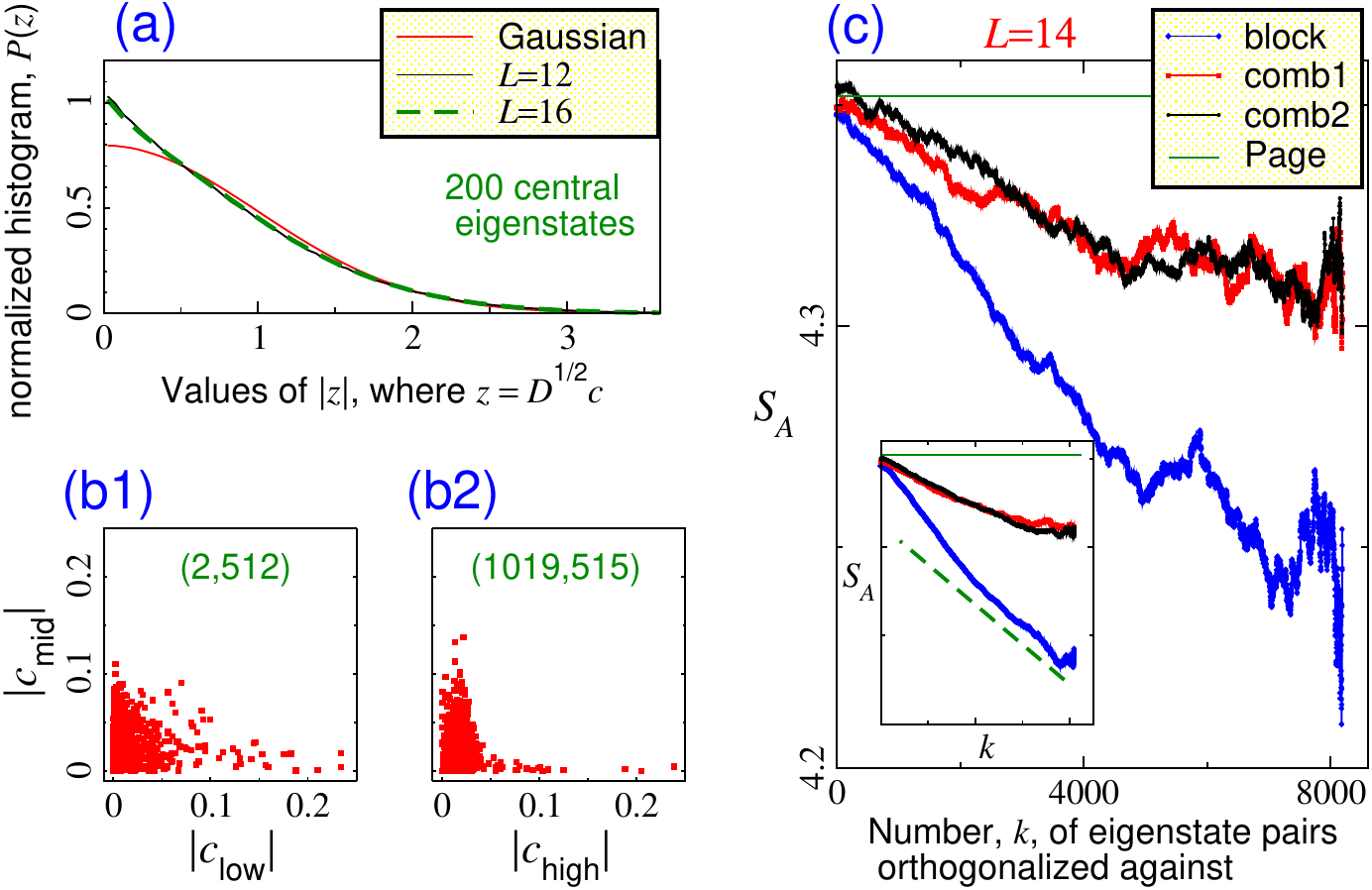}
\caption{Orthogonality blockade, illustrated.
(a)~Eigenstate coefficient distribution, showing an excess of small
  values compared to Gaussian.
Excess appears not to scale with system size.
%
%
(b)~Scatter-plot between eigenstate coefficients.  Each point represents one real-space
configuration.  Coefficients of a mid-spectrum state against those of a low-energy (b1) and a
high-energy (b2) state.  Eigenstates (in brackets) labeled from $1$ to $2^{10}=1024$.
(c)~EE of states obtained by orthogonalizing a random state to $k$ lowest-energy and $k$
highest-energy eigenstates.  $k=0$ is a random state; $k=\frac{D}{2}-1$ is essentially a
mid-spectrum eigenstate.  Inset: EE's averaged over 30 starting states.  Dashed straight line is a
visual guide.
\label{fig:4}
}
\end{figure}

Aspects of the blockade phenomenon are illustrated in Fig.~\ref{fig:4}.
The coefficients of mid-spectrum eigenstates, in the basis of real-space configurations, are not
entirely Gaussian (as would be the case for GOE eigenstates) but have excess weight at small
values \cite{Beugeling_coefficients_PRE2018,LuitzBarlev_PRL16,Baecker2019,Luitz_Khaymovich_BarLev_multifrac_SciPost2020},
Fig.~\ref{fig:4}(a).
Some configurations are ``over-represented'' in eigenstates at the spectral edges; hence by
orthogonality they have to be under-represented in mid-spectrum eigenstates, leading to an excess of
small values of coefficients.  This effect is seen more explicitly when coefficient magnitudes of
different eigenstates are plotted against each other -- Fig.~\ref{fig:4}(b)  shows that
low/high-energy eigenstates  have anomalously large weights in a few basis states, which have small
weights in mid-spectrum eigenstates~\cite{SM}.

In Fig.~\ref{fig:4}(c) we show how orthogonality affects entanglement.  Starting from a random
state, we successively Gram-Schmidt-orthogonalize against pairs of eigenstates at the outermost
edges of the spectrum, i.e., first with the lowest and highest eigenstates, then with the
second-lowest and second-highest eigenstates, and so on.  The EE (both block and comb) of the
resulting states, Fig.~\ref{fig:4}(c), start with zero departure and, after all but the mid-spectrum
eigenstates have been orthogonalized away, end at roughly the values observed for the EE's of
mid-spectrum eigenstates.  The curve has fluctuations depending on the initial random state used,
but the fluctuations decrease with increasing system size and the overall picture is valid for a
variety of random states tried \cite{SM}.  When averaged over various realizations of the starting
random state, one obtains a relatively smooth curve, as shown in the inset.

For simplicity, we have above described the blockade as being due to only the eigenstates at the
spectral edges.  In truth, the mid-spectrum eigenstates are affected by orthogonality to all
non-mid-spectrum eigenstates, not just those at the spectral edges.  Intuitively, one might expect
the spectral edges to have the strongest effect, as these eigenstates are the least generic or
random-like.  Comparing with the straight line in the inset to Fig.~\ref{fig:4}(c), we see that the
slope of the curve is larger for small $k$, i.e., orthogonalizing against the spectral edges has a
stronger effect than orthogonalizing against eigenstates which are intermediate between the spectral
edges and the mid-spectrum region.  This demonstrates that eigenstates closer to the spectral edges
indeed have a stronger role in causing the mid-spectrum departure~\cite{SM}.  For simplicity, we
sometimes loosely describe the blockade to be due to spectral edge states --- it should be
understood that intermediate states also contribute to the phenomenon, albeit to a weaker degree.

Interestingly, the distribution of coefficients by itself does not predict the correct mid-spectrum
EE.  Using a random state with coefficients drawn from the observed mid-spectrum distribution,
Fig.~\ref{fig:4}(a), we find the resulting EE to have a departure one order of magnitude smaller
than that observed in the comb cases.  Thus, a random state with an effective Hilbert space of
reduced dimension $D_{\rm eff}<D$ (as in \cite{DeTomasi_Khaymovich_multifracEE_PRL2020}) is not
sufficient to model the departure.  The orthogonality causes very particular combinations
of configurations to be missing from the mid-spectrum eigenstates; this ``correlated blockade'' is
necessary for the observed departure \cite{SM}.

\section{Rainbow shape implies departure}

The orthogonality mechanism has the following implication.  If the EE versus eigenenergy plot is
rainbow- or arch-shaped, then the correlations in spectral-edge eigenstates which cause those to have low
entanglement will affect the mid-spectrum eigenstates by orthogonality, causing the mid-spectrum
eigenstates to depart from Gaussian randomness.  Thus, a rainbow shape is necessarily accompanied by
a departure in the mid-spectrum states.

We can thus trace back the departure for comb EE to the fact that the EE plot is rainbow-shaped for
comb bipartitioning.  Unlike block partitioning, there is now no parametric argument ($\sim{L^0}$ vs
$\sim{L^1}$) for the rainbow shape, as the spectral edge EE now scales as $\sim{L}$ (the scaling of
the boundary between partitions), the same as the mid-spectrum EE.  A general argument for the
rainbow shape is that, because the low-/high-energy states are more constrained (less like random
states) compared to mid-spectrum eigenstates, the EE at spectral edges has to be farther from the
Page value compared to mid-spectrum EE.  However, this is not obviously related to the spatial
structure of correlations.  In fact, in a model without spatial locality, where all eigenstates have
volume-law scaling, the EE is found to also have a rainbow structure
\cite{Balents_SYK_entanglement_PRB2018}.  According to the picture presented above, orthogonality
should then force a departure of the mid-spectrum EE; indeed this is observed
\cite{Balents_SYK_entanglement_PRB2018}.

\begin{figure}[tbp]
\centering
\hspace{-.5cm}
\centering
\includegraphics[width=\columnwidth]{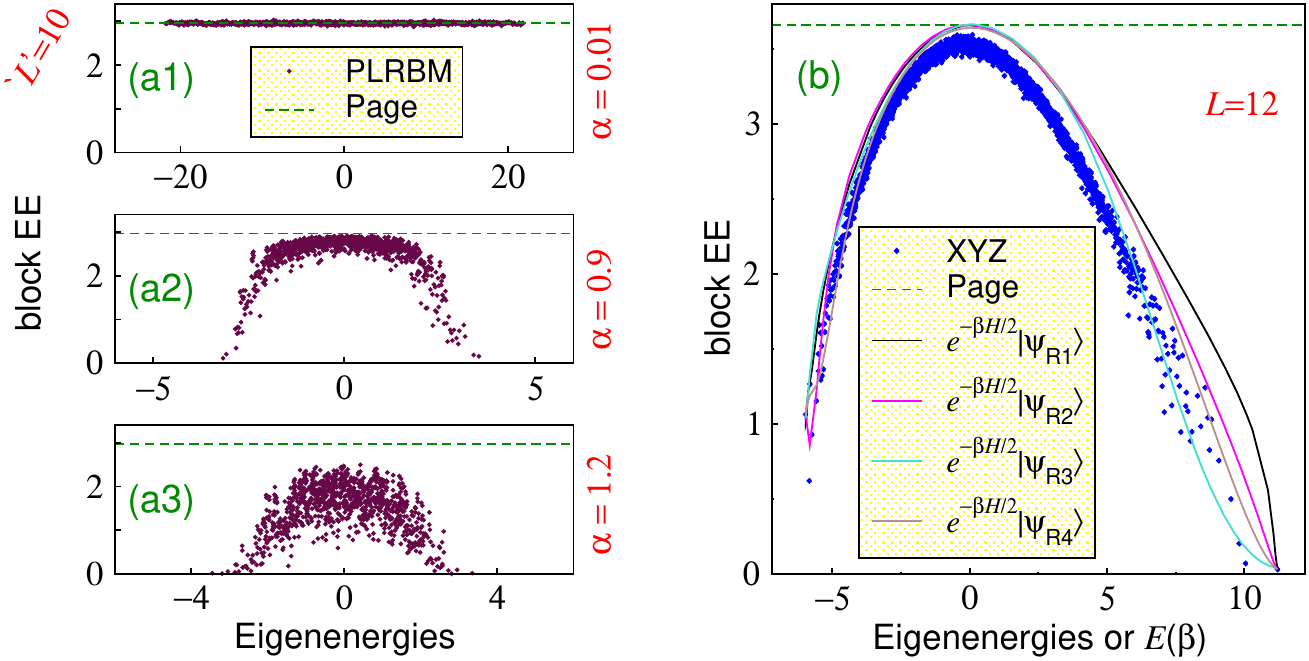}
\caption{
(a)~Power-law random banded matrix of size $2^{10}$, treated as a 10-spin Hamiltonian.  EE versus
eigenvalues, for exponent  $\alpha$ in the ergodic/delocalized (a1), weakly ergodic (a2) and
localized (a3) regimes \cite{SM}.  Case (a2) is analogous to the many-body situation.
(b)~Lines are block EE calculated from states  $e^{-\frac{1}{2}\beta{H}}\ket{\psi_R}$, for different
realizations of random $\ket{\psi_R}$.  On average, these lie above actual eigenstate EE's at all
temperatures/energies.
\label{fig:5}
}
\end{figure}

Our picture is applicable also beyond the context of many-body physics. In power-law random banded
matrices (PLRBMs)~\cite{Mirlin_Fyodorov_etal_PRE1996, Kravtsov_Muttalib_PRL1997,
  Varga_Brown_PRB2000, Mirlin_Evers_PRB2000, Evers_Mirlin_RMP2008,
  Bogomolny_Giraud_entropy_PLBRM_ultrametric_PRL2011,
  Rushkin_Ossipov_Fyodorov_multifrac_critical_JSTAT2011, Mendez_Alcazar_Varga_multifr_EPL2012,
  Bogomolny2018_PLRBM, VegaOliveros_etal_plbrm_PRE2019, Nosov2019correlation,
  DeTomasi_algebraicMBL_PRB2019} and ultrametric
matrices~\cite{Fyodorov_Ossipov_Rodriguez_ultrametric_JSTAT2009,
  Bogomolny_Giraud_entropy_PLBRM_ultrametric_PRL2011,
  Rushkin_Ossipov_Fyodorov_multifrac_critical_JSTAT2011, Mendez_Alcazar_Varga_multifr_EPL2012,
  vonSoosten_Warzel_ultrametric_arXiv2017, Bogomolny2018_PLRBM}, there is a regime of parameters
where the mid-spectrum eigenstates are ``weakly ergodic'' in the sense that, even though the scaling
with matrix size matches the ergodic case, there is deviation from GOE/GUE ensembles at any finite
size~\cite{Bogomolny2018_PLRBM, Nosov2019correlation, Baecker2019}.  To connect to the present
topic: by interpreting the indices of such matrices as spatial configurations~(as discussed in,
e.g.,~\cite{khaymovich2019eth, Lydzba_Vidmar_Rigol_EErandomquadratic_PRL2020,SM}), one can
evaluate entanglements.  For PLRBMs in the weakly ergodic regime, we have found a rainbow-shaped
dependence of EE versus eigenenergy, with a mid-spectrum departure (Fig.~\ref{fig:5}(a)
and~\cite{SM}), just as in the many-body case.  The spectral edges are likely power-law-localized
\cite{SM}, so that the blockade effect is more direct than in the many-body situation.

\section{Context \& consequences}

The EE of non-extremal eigenstates is now the focus of considerable attention, both for chaotic
systems
\cite{Benetti_Casati_PRA2005_entanglement, Deutsch_NJP2010_ThermodynamicEntropy,
  Santos_Polkovnikov_Rigol_PRE2012, Deutsch_Sharma_PRE2013_entropy, Beugeling_entanglement_JSM2015,
  Zhang_Kim_Huse_PRE2015_entanglement, Vidmar_Rigol_PRL2017_EE_chaotic,
  DymarskyLashkariHong2018subsystem, garrison2018SingleEigenstate,
  Watanabe_Sugiura_NatComm2018_VolumeLaw, Fujita_Suguira_Watanabe_pagecurves_JHEP2018,
  DymarskyLashkariHong2018subsystem, Fremling_Haque_NJP2018_LLLdynamics,
  Balents_SYK_entanglement_PRB2018, Murthy_Srednicki_PRE2019, Eisert_entanglementergodic_PRL2019,
  Huang_NPB2019_universal_EE, Huang_Gu_PRD2019_SYK_entanglement, LuGrover2019Renyi,
  Miao_Barthel_eigenstateEE_PRL2021, Morampudi_Chandran_Laumann_constrained_PRL2020,
  Sagawa_HigherOrderETH_PRA2020}
as in this work, and also for integrable systems
\cite{Alba_Fagotti_Calabrese_entanglement_JSTAT2009, Alcaraz_Sierra_cftEE_PRL2011,
  MoelterBarthelSchollwoeckAlba_JSTAT2014, Ares_etal_freefermionEE_JSTAT14,
  Storms_Singh_freefermionEE_PRE2014, KunYang_freefermionEE_PRB2015, Alba_PRB15,
  Beugeling_entanglement_JSM2015, ArnabSenArnabDas_PRB16,
  Vidmar_Hackl_Bianchi_Rigol_freefermionEE_PRL2017, Vidmar_Hackl_Bianchi_Rigol_PRL2018_Ising,
  Zhang_Vidmar_Rigol_trappedfreefermions_PRA2018, Balents_SYK_entanglement_PRB2018,
  CastroAlvaredo_Doyon_PRL2018, Hackl_Vidmar_Rigol_Bianchi_PRB2019_XY,
  LeBlond_Vidmar_Rigol_PRE2019_integrable, Rajabpour_freefermions_HO_EE_PRB2019,
  Barthel_Miao_harmoniclatticeEE_arxiv2019, Lydzba_Vidmar_Rigol_EErandomquadratic_PRL2020}.
The eigenstate EE plays a role in connecting quantum properties to the thermodynamic entropy
\cite{Gemmer_Otte_Mahler_2ndlaw_PRL2001, Deutsch_NJP2010_ThermodynamicEntropy,
  Santos_Polkovnikov_Rigol_PRE2012,  Deutsch_Sharma_PRE2013_entropy,  Storms_Singh_freefermionEE_PRE2014,
Sagawa_fluctuationtheorem_PRL2017,
  garrison2018SingleEigenstate, Murthy_Srednicki_PRE2019}.
Current theory suggests that the mid-spectrum states are effectively random.  We have shown that a
subleading deviation is present for any finite size, and have developed concepts (orthogonality
blockade, residual spatial correlations) pertinent to understanding this deviation.  We expect our
results to be equally valid for systems with symmetries (for which the average random-state EE is
not given by the Page formula), and that the presented concepts will have further
applications, e.g., effects of orthogonality to the extremal eigenstates has been exploited in
recent literature \cite{Nosov2019correlation, Nosov2019mixtures, Kutlin2020_PLE-RG, Deng2020}.  An
open question is whether the deviation \emph{saturates} or vanishes in the large-size limit.

The idea that the state $e^{-\frac{1}{2}\beta{H}}\ket{\psi_R}$ (with $\ket{\psi_R}$ a random state)
is a good model for finite-temperature eigenstates \cite{Sugiura_Shimizu_TPQstates_PRL2012,
  Sugiura_Shimizu_canonical_PRL2013, Watanabe_Sugiura_NatComm2018_VolumeLaw} has been fruitful for
numerical computations of thermodynamic properties \cite{Steinigeweg_Gemmer_Brenig_PRL2014,
  Monnai_Sugita_JPSJ2014, Steinigeweg_Gemmer_Brenig_PRB2015,
  Steinigeweg_HeidrichMeisner_Michielsen_DeRaedt_PRB2015, Steinigeweg_Pollmann_Brenig_PRB2016,
  Hotta_Shimizu_PRL2018, Inoue_eta_sphericalkagome_TPQ_IEEE2019,
  Richter_Steinigeweg_TypicalityPlusLinkedCluster_PRB2019,
  Rousochatzakis_Knolle_Moessner_Perkins_PRB2019, Schnack_Steinigeweg_PRRes2020,
  Nishida_etal_Typicality_Hubbard_JPSJ2020, Schnack_Richter_Steinigeweg_ZNF2020}.
Our observation, that mid-spectrum (highest-EE or infinite-temperature) eigenstates show departures
from random state properties, implies that the state $e^{-\frac{1}{2}\beta{H}}\ket{\psi_R}$ is an
imperfect model for lower-entanglement (finite-temperature) eigenstates as well.  In fact, one can
attempt to reproduce the entanglement rainbow by plotting the EE of the state
$e^{-\frac{1}{2}\beta{H}}\ket{\psi_R}$ against the corresponding energy.  We find that this curve
falls systematically above the EE scatter-plot of actual eigenstates
(Fig.~\ref{fig:5}(b) and~\cite{SM}).


The departure is a signature of deviation from GOE/GUE behavior at all finite sizes.  Signatures of
this deviation also appear in eigenstate coefficient distributions (Fig.~\ref{fig:4}(a); also
\cite{Beugeling_coefficients_PRE2018, LuitzBarlev_PRL16, Baecker2019,
  Luitz_Khaymovich_BarLev_multifrac_SciPost2020}).  Mid-spectrum eigenstates have the same scaling
behavior as GOE/GUE, but approach the thermodynamic limit differently, as also seen in
multifractality analysis \cite{Baecker2019}.  This behavior could justifiably be called ``weakly
ergodic'', although the phrase does not yet have a widely accepted definition
\cite{Bogomolny2018_PLRBM, Nosov2019correlation,
  Nosov2019mixtures,Baecker2019,DeTomasi_Scardicchio_Khaymovich_PRB2020,
  Khaymovich_Kravtsov_Altshuler_Ioffe_PRRes2020}.  In
Ref.~\cite{Khaymovich_Kravtsov_Altshuler_Ioffe_PRRes2020}, weak ergodicity is associated with
breaking of ``basis rotation invariance.''  Our finding, that mid-spectrum eigenstates have
different block EE and comb EE, is another type of non-invariance under basis rotation.

Note added in Proof: After this paper was accepted, we became aware
of Ref.~\cite{HUANG}, which quantitatively accounts for part of the departure
for block partitions, in the case where the Hamiltonian is local.

\begin{acknowledgments}
We thank A.~B{\"a}cker and G.~De~Tomasi for useful discussions.
The work of I.M.K. is supported by the Russian Science Foundation under the
grant 21-12-00409.
\end{acknowledgments}

\bibliography{references}

\end{document}